\documentclass[conference]{csce}
\usepackage{comment}
\usepackage[hmargin=.75in,vmargin=1in]{geometry}
\usepackage[american]{babel}
\usepackage[T1]{fontenc}
\usepackage{times}
\usepackage{algorithm}
\usepackage{algorithmicx}
\usepackage{algpseudocode}	 
\usepackage{mathrsfs}
\usepackage{bm}
\usepackage{color}
\usepackage{xcolor}
\usepackage{pdflscape}
\usepackage{graphicx} 
\usepackage{float}
\usepackage{algorithm}
\usepackage{algorithmicx}
\usepackage{algpseudocode}	
\usepackage{indentfirst}
\usepackage{slashbox}
\usepackage{booktabs}       
\usepackage{amsfonts}       
\usepackage{nicefrac}       
\usepackage{microtype}      
\usepackage{lipsum}
\usepackage{setspace}
\usepackage{array}
\usepackage{comment}
\usepackage[T1]{fontenc}    
\usepackage{hyperref}       
\usepackage{url}            
\usepackage{booktabs}       
\usepackage{amsfonts}       
\usepackage{nicefrac}       
\usepackage{microtype}      
\usepackage{lipsum}
\usepackage{xspace}
\usepackage{times}
\usepackage{amsmath}
\usepackage{booktabs}
\newcommand\MyBox[2]{

  \fbox{\lower0.75cm
    \vbox to 1.75cm{\vfil
      \hbox to 1.75cm{\hfil\parbox{1.4cm}{#1\\#2}\hfil}
      \vfil}%
  }%
}
\AtBeginDocument{%
  \providecommand\BibTeX{{%
    \normalfont B\kern-0.5em{\scshape i\kern-0.25em b}\kern-0.8em\TeX}}}
\usepackage{flushend}

\usepackage{textcomp}
\usepackage{epsfig,graphicx}
\usepackage{xcolor}
\usepackage{amsfonts,amsmath,amssymb}
\usepackage{booktabs}

\title{ \bf Malware Detection using\\ Artificial Bee Colony Algorithm
}
\author{Farid Ghareh Mohammadi$^1$, Farzan Shenavarmasouleh$^1$,  M.  Hadi Amini$^2$,and Hamid R.  Arabnia$^1$\vspace{0.15in}\\
1:  Department of Computer Science,  Franklin College of Arts and Sciences \\ University of Georgia,  Athens,  Georgia,  30602  \\
2:  School of Computing and Information Sciences,  College of Engineering and Computing\\Sustainability, Optimization, and Learning for InterDependent networks laboratory (solid lab)  \\Florida International University,  Miami,  FL 33199 \\
Emails :  farid.ghm@uga.edu, farzan.shenavarmasouleh@uga.edu,  moamini@fiu.edu, hra@uga.edu}

\begin{document}
\newcolumntype{M}[1]{>{\centering\arraybackslash}m{#1}}
\newcolumntype{N}{@{}m{0pt}@{}}

\maketitle                        

\begin{abstract}
  Malware detection has become a challenging task due to the increase in the number of malware families. Universal malware detection algorithms that can detect all the malware families are needed to make the whole process feasible. However, the more universal an algorithm is, the higher number of feature dimensions it needs to work with, and that inevitably causes the emerging problem of Curse of Dimensionality (CoD). Besides, it is also difficult to make this solution work due to the real-time behavior of malware analysis. In this paper, we address this problem and aim to propose a feature selection based malware detection algorithm using an evolutionary algorithm that is referred to as Artificial Bee Colony (ABC). The proposed algorithm enables researchers to decrease the feature dimension and as a result, boost the process of malware detection. The experimental results reveal that the proposed method outperforms the state-of-the-art.
\end{abstract}

\vspace{1em}

\section{Introduction}
 \begin{algorithm*}
\begin{minipage}{1.\linewidth}
	\caption{Implementation of ABC algorithm for feature selection inspired by \cite{mohammadi2014image}}
	\begin{algorithmic}
		\Require  $S=\{x_0,  x_1, x_2, ..., x_n\}$,  $limit  \ge 0 $,  $ 0 \le \text{lower Bound} \le n/2$,   $\text{lower Bound} \le \text{upper Bound}  \le n$, $ max_{iteration} \ge 0$,   t=0 ,  $Best {Solution} = \emptyset$.
		
		\Ensure \texttt{Best {Solution} :} $F=\{x_0, x_1, x_2, ..., x_m\}$ ,   m$\le$ n , $({\forall f_i \in F})\in S$ ,  $ F_{length} \le  S_{length}$.
    \\
    	\State {Evaluate the whole food source using SVM}
    		\Comment{\textbf{Step 1:} Explore L dimension using Employed bees}
		\For \{t=0 $\cdots$ $max_{iteration}$\}

		\State Exploit the local foods to generate new food 
		\Comment{\textbf{Step 2:} Exploiting by using Onlooker bees}
		 \State Choose parents and generate  a new food (solution) based on  ${V_i}$= ${f_i}$+${v * (f_i-f_j)}$
		 \If{limit is met} : 
	     	\State{\{ }\State{Explore new (unseen) food source to prevent from local optimum}
	     	\Comment{\textbf{Step 3:} Exploring by using Scout bee}
	     	\State${X_i}$= ${X_{\text{ upper Bound}}}$+${v'}$ * \{ (${X_{{\text{ upper Bound}}}}$-${X_{\text{lower Bound}}}$)\}\State {\}}
	     	
		 	\Return ${New Solution}$
	\EndIf
		\State Call fitness function to evaluate the Solution 
	   \If{ any Solution obtained the best score}
	   \State \hspace{3mm} $\{\text{Update the Best}\text{Solution}\}$
	   \EndIf
	   		\EndFor
	\end{algorithmic}
	\label{FSABC}
\end{minipage}
\end{algorithm*} 
\subsection{Motivation}
 Traditional machine learning algorithms have always been the building blocks of malware detection systems in the last few decades and great successes have been achieved by them. However, the majority of these systems leverage supervised learning strategies to detect malwares, and hence they heavily rely on formerly seen and labeled static features. In many cases, these features may not be relevant enough to enable researchers to obtain promising results and use the algorithms to their full potential. In this paper, we aim to propose a new malware detection system that employs an evolutionary algorithm and is capable of selecting the most important and the most relevant features that are correlated with the seen and labeled data. The evolutionary algorithm that gives us a hand in the process of feature selection is referred to as Artificial Bee Colony (ABC).
  
  ABC algorithm is developed by Karaboga \cite{karaboga2005idea} and provides convincing solutions for continuous and discrete optimization problems. It is considered as one of the most powerful optimization algorithms since it simulates the foraging behavior of honeybees that helps them find the best food source available \cite{ch1_farid}. ABC works based on three types of bees, namely, employed bees, onlooker bees and, one scout bee. 
The process of feature selection gets started with the employed bees computing the accuracy of each solution. Then the onlooker bees seek the global maximum solution according to the best accuracy, named as crossover sub-process, which creates a new offspring out of selected high solutions to create a higher accuracy one. Onlooker bees run until the limit condition is met. Next, one of the employed bees converts into a scout. The scout bee explores beyond the boundary in which the onlookers seek the global maximum accuracy solution. Note that, in each iteration, Scout bee behave like a mutation process, which is the final phase of normal evolutionary algorithms. We provide more details of the work in the proposed method. This entire process repeats itself until no higher maximum accuracy solution can be found or the number of crossover iterations reaches the pre-designed Limit \cite{ch2_farid}.  

The works in the state-of-the-art undeniably prove the importance of feature selection \cite{fang2019feature, fatima2019android}. The new emerging trend in the feature selection process is to leverage evolutionary algorithms. Fatima \emph{et al} \cite{fatima2019android} presented a new algorithm to employ genetic algorithms in the feature selection phase to improve the accuracy of malware detection. We choose ABC since it yields a better result than the genetic algorithm. Furthermore, the number of parameters that it uses is much less than the genetic algorithm.
\subsection{Related work}
Yerima and Sezer \cite{yerima2018droidfusion} presented DroidFusion, which is a process of supervised classification. It comprises two levels. In the first level, several low-level classifiers like Random Tree, J48, and more are trained and ensembled together. Then, in the second level, a set of ranking-based algorithms are applied to the generated models to generate the final one.
However, the weakness of this paper is that the authors did not make use of better classifiers like support vector machine (SVM).  Also, they did not remove unrelated features that significantly impact the accuracy of the final result. 
Wang \emph{et al} \cite{wang2018detecting} adopted different static feature categories in an SVM classifier for filtering and then fed them as an input to an ensembled classifier consisting of five models, namely, Classification and Regression Tree (CART), Random Forest, Naive Bayes, K-Nearest Neighbors, and SVM. This research study still suffers from the curse of dimensionality problem \cite{ch1_farid}. Consequently, one of the solutions is to use an evolutionary algorithm for feature selection \cite{ch2_farid}.

Bazrafshan \emph{et al} \cite{bazrafshan2013survey} provided an overview of challenges in the malware detection methods, talked about how these shortcomings are covered alternatively, and presented three important techniques employed to defeat them. One signature-based, one heuristic-based, and one behavior-based method. The goal of the paper is to develop malware detection techniques using heuristic and evolutionary algorithms. It is noteworthy that the authors studied some key features for applications toward malware detection that enable the malware to skip the detection. So this result expresses that finding the important relevant features is promising in the process of malware detection. Further, Bradicich \emph{et al} \cite{bradicich2012heuristic} presented a heuristic-based malware detection system that uses a base-line inventory of file attributes per each associated file and finally detects the irrelevant features and make the features united. This work also proves the importance of feature selection. In addition to them, researchers \cite{green2015detecting} further explored the malware detection process using statistics as features which are accumulated as part of processing the collection, with tuned heuristic analysis iteratively.

Fatima \emph{et al} \cite{fatima2019android} presented an algorithm for malware detection using feature selection. The authors applied evolutionary GA to select the best features by decreasing the feature dimensionality of the original dataset, so that they could train the classifier to detect the malwares. They used two classifiers: one SVM and one neural network for that end.

\subsection{Contribution}
Finding the optimal data and leveraging them to generate a model that yields a high performance is challenging \cite{ch1_farid,ch2_farid}. In the world of malware detection, there are several research studies done to apply evolutionary algorithms to deal with this challenge \cite{fang2019feature}. To that end, we find the artificial bee colony the most robust and time-efficient evolutionary algorithm that has the work done expeditiously \cite{ch2_farid, bazrafshan2013survey}. Our main contributions are as follows:
 
 \hspace{20pt}$\diamond$ \textbf{First:} The artificial bee colony algorithm operates in continuous space by default. Hence, as our first contribution, we modify the innate continuous behavior of the artificial bee colony algorithm and convert it into a discrete algorithm.
 
 \hspace{20pt}$\diamond$ \textbf{Second:} We present a new feature selection algorithm using the artificial bee colony algorithm to improve the overall malware detection performance.
\par

\subsection{Organization}
The rest of this paper is organized as follows. In section 2, we present an overview of the classic machine learning algorithms, their preliminaries, and their learning process. Next, the Intersection of machine learning and medical sciences will be discussed in section 3. Then, in section 4, we provide comprehensive information about deep learning algorithms. Finally, in section 5, we discuss the applications of deep learning for infectious diseases followed by assessment criteria.

\section{Proposed Method}
In order to contribute to this field and improve the state-of-the-art while extending them properly for work reusability as discussed in \cite{farzan9071340}, we propose a feature selection based malware detection system by extending the work presented in \cite{mohammadi2014image}. In this system, we aim to propose a discrete version of the artificial bee colony (ABC) algorithm to perform the task of feature selection as shown in Algorithm \ref{FSABC}. We choose ABC since it works very accurately for the task of image classification \cite{mohammadi2014image,mohammadi2017region}. Hence we hypothesize that it can perform well for malware detection when customized adequately. Our Algorithm performs feature selection process in three main steps as follows:

\textbf{Step 1:} It explores the whole food source to recognize the possible good features by calculating the accuracy of each food using employed bees. The employed bees start by taking each food from the food source and calculate the nectar (goodness) or accuracy by using a machine learning algorithm called support vector machine (SVM).

\textbf{Step 2:} Next, it aims to exploit each food source and find the best possible food sources to generate a new food source using onlooker bees. These bees choose new parents and generate new food sources using the following formula:

{
\par \hspace{50pt} ${V_i}$= ${f_i}$+${v * (f_i-f_j)} $ \hspace{10pt}  (1) \par
}

where, i,j={0,1,2…, N},  N=upper bound of the features, $f_i$ stands for the goodness (accuracy) of the feature assigned to the employed bee, $f_j$ stands for the goodness of the feature the onlooker has chosen, v stands for a random real number in the range of -1 and +1, and $V_i$ is the goodness (accuracy) of the newly generated solution that has been selected by the onlooker Bee. If $V_i<f_i$ we conclude that the goodness of $f_j$ is more than  ${V_i}$, and hence we have to replace j with i in order to improve the total goodness (accuracy) of the food source. But if $V_i$>$f_i$, then we continue to the next iteration.

\begin{table*}[ht]
\center
  \caption{Evaluation criteria}
  \vspace{0.1in}
  \renewcommand{\arraystretch} {1}
  \begin{tabular}{|M{5cm}|M{5cm}|N}
    \hline 
    \textbf{Assessment Criteria} &{Formula}  \\ 
             \hline
      Accuracy (ACC) &\vspace{8pt} $ \frac{ TP + TN}{TP+FP+TN+FN}$\\[10pt]
      \hline
       Specificity & \vspace{8pt} $ \frac{ TN}{TN+FP}$ \\[10pt] \hline
       Recall or sensitivity &\vspace{8pt}$ \frac{ TP}{TP+FN}$\\[10pt] \hline

    \end{tabular}
       \label{tab:assessmentCriteria}
  \end{table*}
\textbf{Step 3:} In the final step which is considered as the mutation step of evolutionary algorithms, it uses a scout bee to produce a food source randomly through a predefined search space (original feature dimension). Once the goodness of the recent food source has not been enhanced by the predefined number of iterations, referred to as limit, then, the food source will be abandoned. After that, the scout bees will randomly try to identify a novel food source position in the original feature dimension calculated by Eq. (2). This process is performed to prevent getting stuck in a local optimum.

\par ${X_i}$= ${X_{\text{ upper Bound}}}$+${v'}$ * \{ (${X_{{\text{ upper Bound}}}}$-${X_{\text{lower Bound}}}$)\}   \hspace{10pt}   (2) \par

where $X_{max}$ and $X_{min}$ are the upper and lower bounds of the number of population (or the original feature dimension) respectively and v\rq is chosen randomly from the range of 0 to 1.

\textbf{Termination condition:} The aforementioned phases of ABC algorithm will be repeatedly run, and the employed, onlooker, and scout bees will keep doing their duties until either the number of runs reaches the max iteration count or the condition of best food source is satisfied which is the highest accuracy of detecting malwares.

  \begin{figure*}[ht]
    \center
    \includegraphics[height=2in]{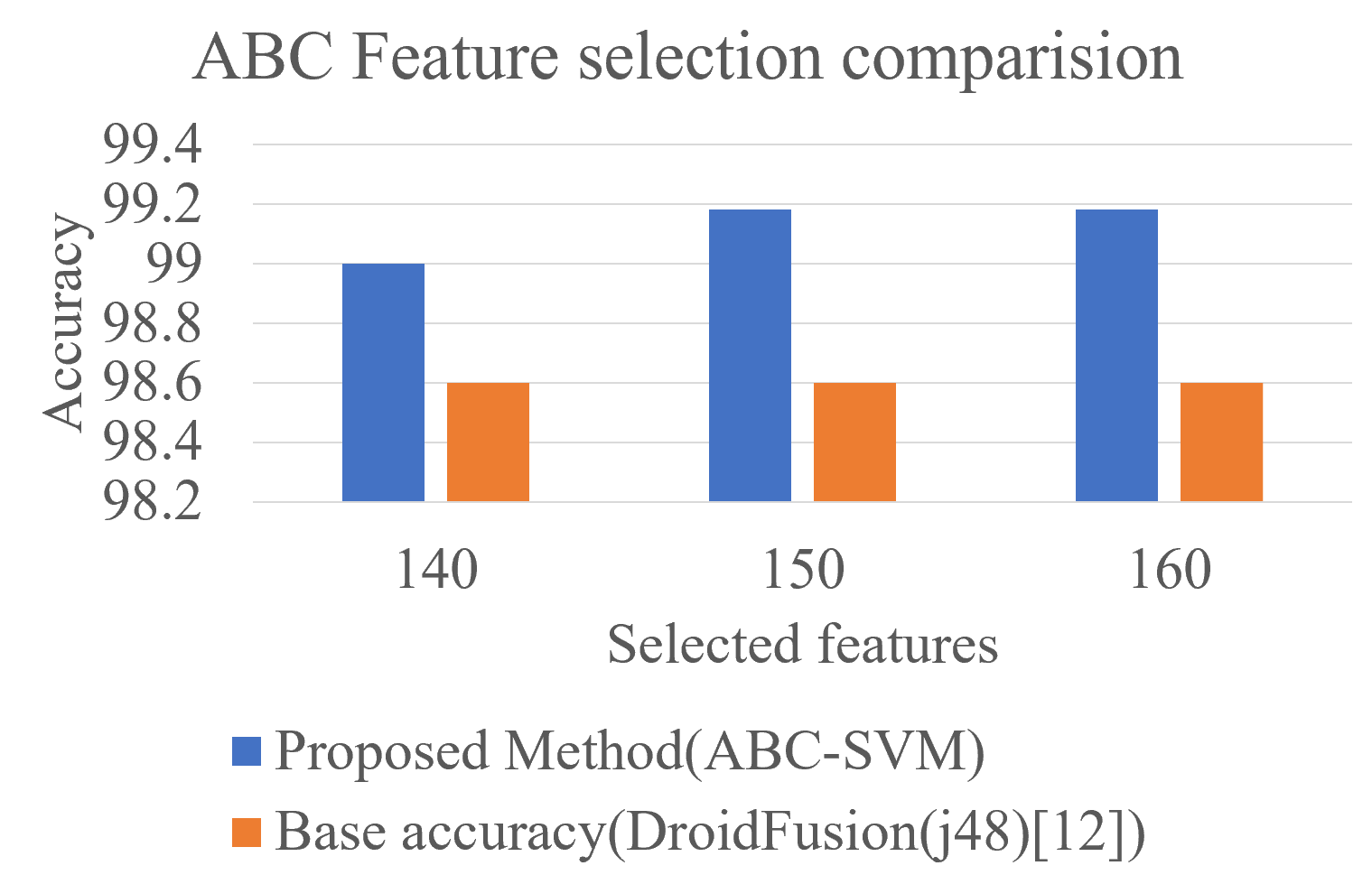}
    \caption{Feature selection comparison}
    \label{fig:FS}
\end{figure*}

\begin{table*}[ht] 
\center
  \caption{Comparing the related literature with our study(\checkmark:considered; $\times$ : not considered)}
  \vspace{0.5cm}
  \renewcommand{\arraystretch} {1}
  \hspace{-1cm}\begin{tabular}{|c|c|c|c|c|}
    \hline 
    \backslashbox{\textbf{Approach}}{\textbf{Criteria}}  & Recall ($\%$) &	Specificity($\%$) & Accuracy ($\%$)    \\ 
         \hline
       DroidFusion (J48) \cite{yerima2018droidfusion}  &	98.4 &	998.9 & 98.6 \\ \hline
      
       Proposed method (ABC+SVM)  & \textbf{98.9} &	\textbf{99.46} & \textbf{99.18}  \\ \hline
    \end{tabular}
      \label{comparisonResult}
  \end{table*}
\section{Experimental result}
\subsection{Dataset}
The dataset that we use in this study was called DERBIN, and included 215 feature dimensions which were extracted from 3799 applications; 1260 of which were malware apps and the rest were benign apps. The dataset in \cite{yerima2018droidfusion} has been extracted and used to design and implement a multilevel classifier fusion approach for the aim of Android malware detection. This dataset is available online with a supporting file describing each feature collected through static code analysis of the Android apps.

\subsection{Evaluation criteria}
We use three prevailing criteria, namely, sensitivity or recall, specificity, and accuracy, to carefully evaluate our proposed method and properly compare it to the state-of-the-art \cite{farzan9071340} \cite{yerima2018droidfusion}.

\subsubsection{The importance of usage}
{\textcolor{black}{Assessing the goodness of results generated by the models is highly domain-specific. Estimating the ratio of correctly classified malwares as malicious applications to the total number of malwares within the dataset, i.e including true positive and false negative, is referred to as recall or sensitivity. This can reveal how sensitive a built malware detection system is while facing new test data. The higher the recall, the more sensitive it is. Using recall, we look for the positive samples of data rather than the other type which is benign. The reason is that it is crucial for us to prevent detecting a real malicious application as a benign, but detecting a benign application as a malware does hurt the system. If the value of recall is not high, hosts which can be smartphones or personal computers would be vulnerable to get hacked and get their data stolen.}}
{\textcolor{black}{
Additionally, we compute specificity, which is the ratio of the number of correctly detected benign samples to all of the benign samples. The specificity of the model illustrates how robust and reliable the generated model is. If the specificity of the model is low, the host would get hurt by incidents like malfunctioning or rebooting problems due to lack of the important files which had already been quarantined or deleted by the malware system.}}

{\textcolor{black}{
Furthermore, accuracy is one of the most common criteria in all domains, especially in the malware detection system. Calculating accuracy solely gives us the performance of the malware detection system in general rather than the performance of it on detecting each class. By having the accuracy, we cannot decide whether the generated model is reliable or not as we are not aware of the system failures, i.e which class the system fails more on (malicious or benign).
 }}

\subsection{Results}
In this study, we apply the proposed method on the DERBIN dataset and calculate the evaluation criteria. First, We evaluate our method based on accuracy to identify the most important features of the DERBIN dataset. We examine several sets of features such as 90, 100, 110, 120, 130, 140, 150, and 160. We realize that 140, 150, and 160 were the most important sets of features. Anything less than 140 negatively affected the accuracy and anything above 160 was redundant, because increasing the number of features did not help to improve accuracy. Also, The accuracy of the set of 150 features was identical with the set of 160 features. Thus, we choose the former due to less number of features.

Figure \ref{fig:FS} illustrates the different sets of features that are selected with the associated accuracy in comparison with the base accuracy presented in \cite{yerima2018droidfusion}. This figure shows that the best set of features is 150 and we choose this set to calculate the final results comprising recall, sensitivity, and accuracy of the method. Table \ref{comparisonResult} reveals the experimental result with the evaluation criteria.

\section{Conclusion}
The main problem almost in all domains is the curse of dimensionality, and the malware system world is no exception. Detecting malwares in real-time plays a significant role in having robust and safe applications on smart devices and preventing data theft. In this paper, we address both issues by proposing an evolutionary-based malware detection system to improve the accuracy of detection. We employ the artificial bee colony algorithm as a simple and time-efficient evolutionary algorithm to select the most relevant features. {\textcolor{black}{The experimental results reveal that our proposed method yields recall, specificity, and accuracy of 98.9\%, 99.46\%, and 99.18\% respectively. These illustrate that the proposed method works expeditiously due to selecting the most important features and performing feature dimension reduction.}}

\bibliographystyle{IEEEtran}
\bibliography{bib.bib}
\end{document}